\documentclass[
final            
  ]
  {aipproc}

\layoutstyle{8x11double}

\usepackage{graphicx}
\usepackage{dcolumn}
\usepackage{bm}
\usepackage{amssymb}
\usepackage{amsmath}
\begin{document}

\title{Participation And performance In 8.02x Electricity And Magnetism: The First Physics MOOC From MITx}


\author{Saif Rayyan$^{1}$, Daniel T. Seaton$^{2}$, John Belcher$^{1}$,\\David E. Pritchard$^{1,3}$, Isaac Chuang$^{1-4}$}{
  address={Department of Physics$^{1}$, Office of Digital Learning$^{2}$, Research Laboratory for Electronics$^{3}$, Electrical Engineering and Computer Science$^{4}$, Massachusetts Institute of Technology, Cambridge, MA 02139.}
}

\begin{abstract}
Massive Open Online Courses are an exciting new avenue for instruction and research, yet they are full of unknowns. In the Spring of 2013, MITx released its first introductory physics MOOC through the edX platform, generating a total enrollment of 43,000 students from around the world. 
We describe the population of participants in terms of their age, gender, level of education, and country of origin, highlighting
both the diversity of 8.02x enrollees as well as gender gap and retention. Using three midterm exams and the final as waypoints, we highlight performance by different demographic 
subpopulations and their retention rates. Our work is generally aimed at making a bridge between available MOOC data and topics associated with the Physics Education Research community. 
\end{abstract}

\maketitle

\section{Introduction}

\hspace{0.19in}Massive Open Online Courses (MOOCs) are an exciting, yet unknown, new model for instruction. At the time of writing this article, hundreds of courses from a few dozen universities are available through a handful of providers (e.g. Coursera, edX, and Udacity). The majority of current MOOCs are affiliated with top universities from around the world, generating excitement regarding how these free courses will impact higher education, as well as enthusiasm to worldwide participants who desire access to courses from top universities. These courses also provide an exciting avenue for understanding how students learn through analysis of their logged interactions.

Physics education research provides a perspective concerned with evaluating pedagogies and classroom practices through comparison of students scores and standardized assessments across multiple classes, as well as by running semi-controlled experiments within single classes of tens to hundreds of students. MOOCs provide an intriguing opportunity to study large student populations all within the setting of a single course, potentially providing great statistics (e.g. around 8000 participants finished 6.002x - the inaugural edX course~\cite{SeatonMOOC2013}). However, challenges emerge when considering the diversity of enrollment and variation in learner backgrounds~\cite{breslow2013mooc}. In order to use MOOCs as a laboratory for Physics Education Research, one needs to first understand this diversity and how it persists through a given course.

This study is a preliminary analysis of the first edX physics MOOC: 8.02x- Electricity and Magnetism. We explore the diversity of the student population of 8.02x through self-reported demographics, highlighting age, gender, country of origin and level of education. We further explore retention from initial enrollment to the final exam. As a way of exploring how diversity affects course outcomes, we analyze retention and performance on exams in the context of the different demographic groups.

\section{Course Description}
\hspace{0.19in}8.02x is a Massive Open Online Course offered through edX in Spring of 2013. The course is centered around recorded video lectures of Introductory Physics II (8.02) taught by MIT's Walter Lewin in 2002. To promote engagement and provide self-assessment, lectures are broken into roughly 5-10 segments, with machine graded multiple choice questions interspersed (worth 5\% of the grade). Other components of the course include machine graded weekly homework sets (18\%) , interacting with simulations (developed for TEAL~\cite{Dori2004}) and answering related concept questions (2\%), videotaped problem solving sessions, a complete e-Textbook, three midterm exams (worth 15\% each) and a final exam (worth 30\%). A threaded discussion board enabled students to interact with each other, and also provided limited interaction with course staff. Certification is granted to students whose scores are greater than 60\%. The course ran for 16 weeks from February 2013 to June 2013, equivalent in length to a semester long course at MIT. 

\begin{figure*}
\includegraphics[width=2\columnwidth]{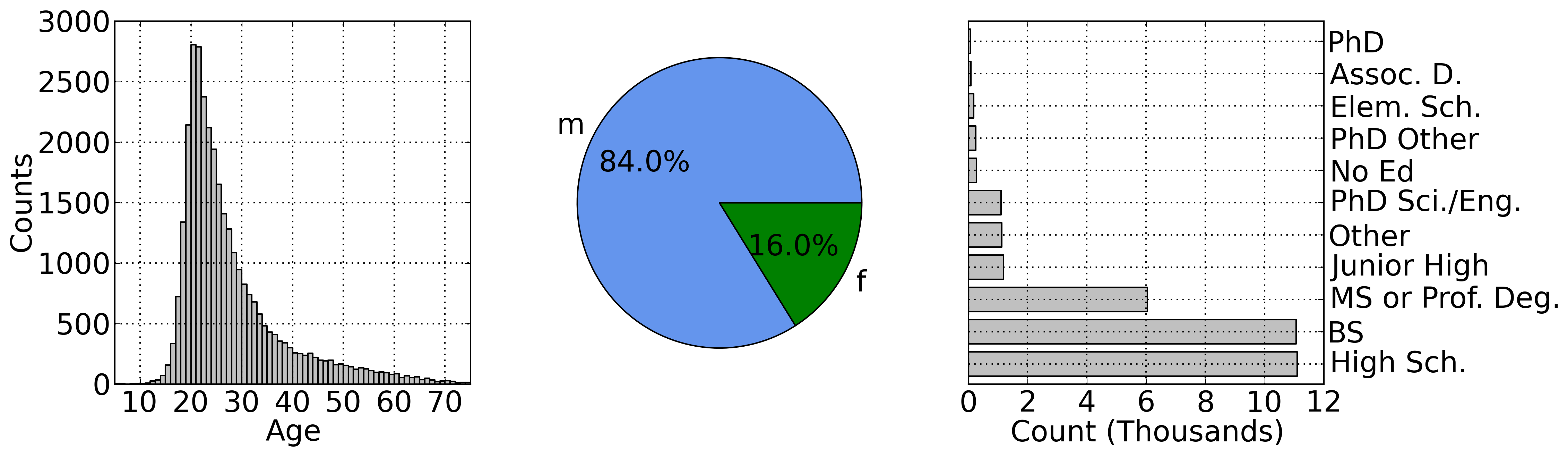}
\caption{\label{fig:dem} 
Self-reported demographics for enrollees in the Spring 2013 instance of 8.02x; Age, Gender, and Level of Education. 
}
\end{figure*}

\section{Data and Methodology}
\hspace{0.19in}Three types of data collected by the edX platform are analyzed: self-reported demographics provided at enrollment, performance metrics generated by the instructor dashboard, and limited click-stream data stored in tracking-logs. The majority of our analysis is centered around the performance metrics for examinations, which were obtained through the instructor dashboard and was fully available for this analysis. Because anyone can log-in and view an examination (while they are open), participants are only counted if receiving credit higher than zero. Due to the large weights given to the examinations, we view participation as a proxy for overall course participation. This is reasonable considering how certificate earners behave over the course of a MOOC~\cite{SeatonMOOC2013}, but does leave out small populations using MOOCs in non-traditional ways~\cite{Kizilcec2013}. 

Click-stream data provide a measure of overall activity of each participant. Since there is no penalty for browsing course content, particularly considering the press surrounding MOOCS, it is expected that many people simply register and log in to browse content, without ever intending to take the course. Click-stream data are used to indicate activity levels of participants, providing an estimate of the number of people active in the course before the first midterm (At the time of this article, we only had access to click-stream data for roughly two-thirds of the course). Participants with greater than 500 events before the first midterm are considered active participants (one event is equivalent to playing a lecture video, opening a book page, etc., and 500 events are equivalent to the average activity over two of the four week proceeding the first midterm).


\begin{table}
\caption{\label{tab:cc}Selected demographics against country of origin: enrollment count, mean age, percentage of enrollment older than 25, and percentage females.}
\begin{tabular}{ @{\extracolsep{\fill} } lrrrr}
    		& & & & \\
		& Enroll. & Mean & \% Age &  \% Female  \\
		& Count	 & Age & > 25 & \\ \hline
    All 			& 43758 	 & 27.5 	& 31.6 	& 16.0	\\
    United Sates 	& 5670	 & 32.7 	& 40.7 	& 17.2	\\
    India 			& 5567	 & 22.6 	& 12.1 	& 13.9	\\
    Nigeria 		& 1265	 & 28.2 	& 19.7 	& 17.7 	\\
    Brazil 			& 1047	 & 27.1 	& 32.0 	& 13.8	\\ 
    United Kingdom 	& 1006	 & 31.0 	& 39.2 	& 15.3	\\ \hline
    \end{tabular}
\end{table}



\subsection{Who enrolled in 8.02x?}
\hspace{0.19in}8.02x enrolled 43,758 participants from around the world, of which, 32,504 provided responses to a demographic survey given at the time they sign up for an account on http://www.edx.org . Self-reported demographics include age, gender, highest level-of-education attained, country of origin, and their language. Figure~\ref{fig:dem} provides a summary of demographics for the 8.02x enrollee population. The distribution of age (Left) is skewed toward college-aged students and has a long tail extending to higher ages. The mean, median, and mode of the age distribution are 27.5, 28, and 20, respectively. The distribution of gender (Middle) is heavily shifted toward males (84\% of the population). The males to females ratio is similar to physics majors graduating from U.S. bachelor's programs~\cite{Mulvey2012}. The level of education attained by enrollees consists mainly of High School ($\approx34\%$), Bachelor's of Science ($\approx34\%$),  Master's or Professional Degree ($\approx19\%$), Junior High School ($\approx3.6\%$) and Science/Engineering PhD ($\approx3.4\%$).

Participants originating from USA and India represent the two largest groups, followed by Nigeria, Brazil and United Kingdom. Table \ref{tab:cc} highlights differences in the age distribution between participants from the US and participants from India. The mean age of participants from the US is 32.7, with $\approx$ 40\% of the population older than 25. In contrast, India has a lower average age of 22.7, with only 12\% older than 25. Similar trends were observed in a smaller physics MOOC ~\cite{Fredericks2013} , and other MOOC offerings outside physics ~\cite{Belanger2013}.

\begin{figure}
\includegraphics[width=0.95\columnwidth]{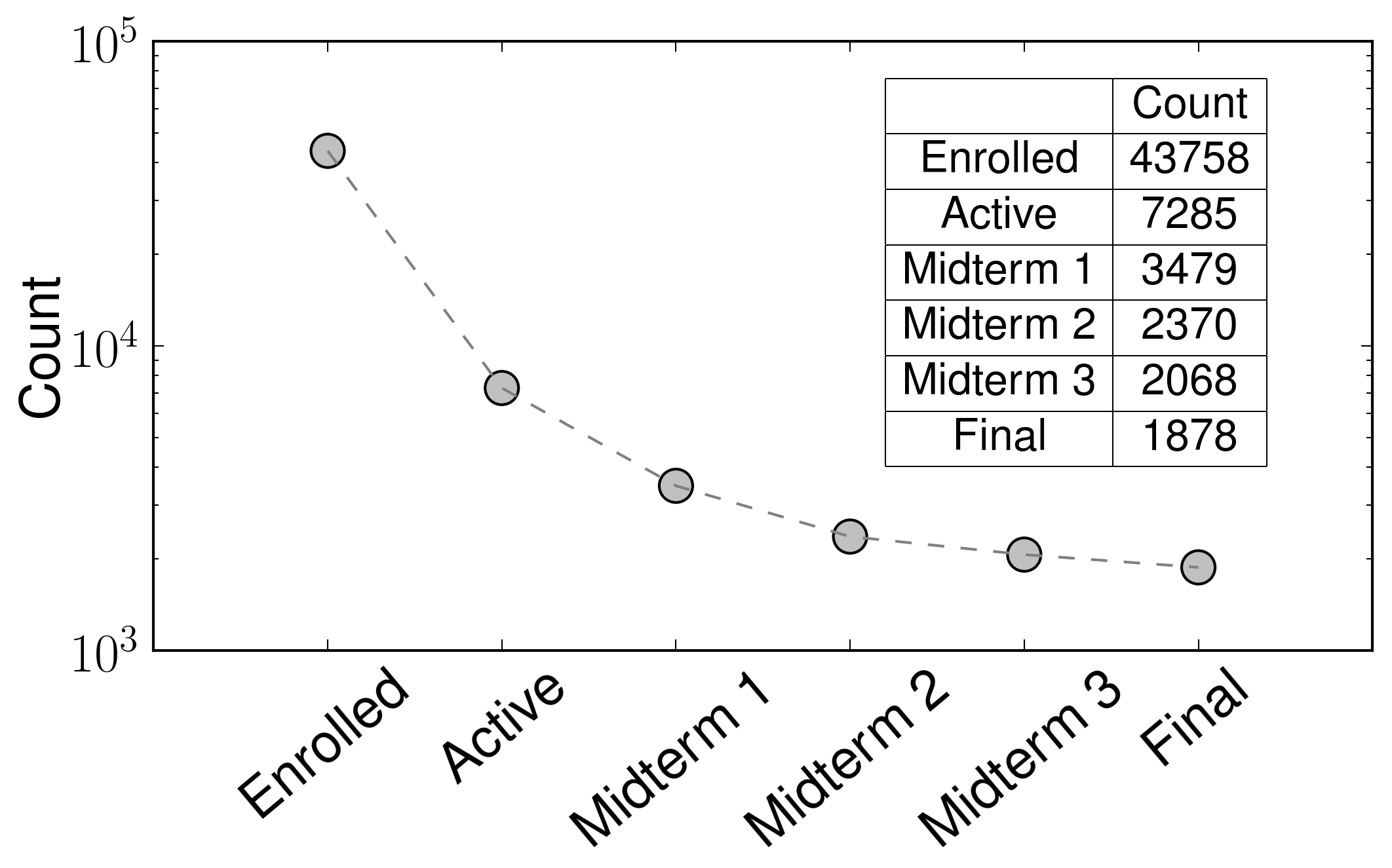}
\caption{\label{fig:retention} 
Plot of number of enrollees, active students, and participants in examinations. Enrollees join at any time, while Active participants are counted over the first four weeks of the course, and examinations proceed in a linear fashion spaced roughly three weeks apart.
} 
\end{figure}

\begin{figure}
\includegraphics[width=0.975\columnwidth]{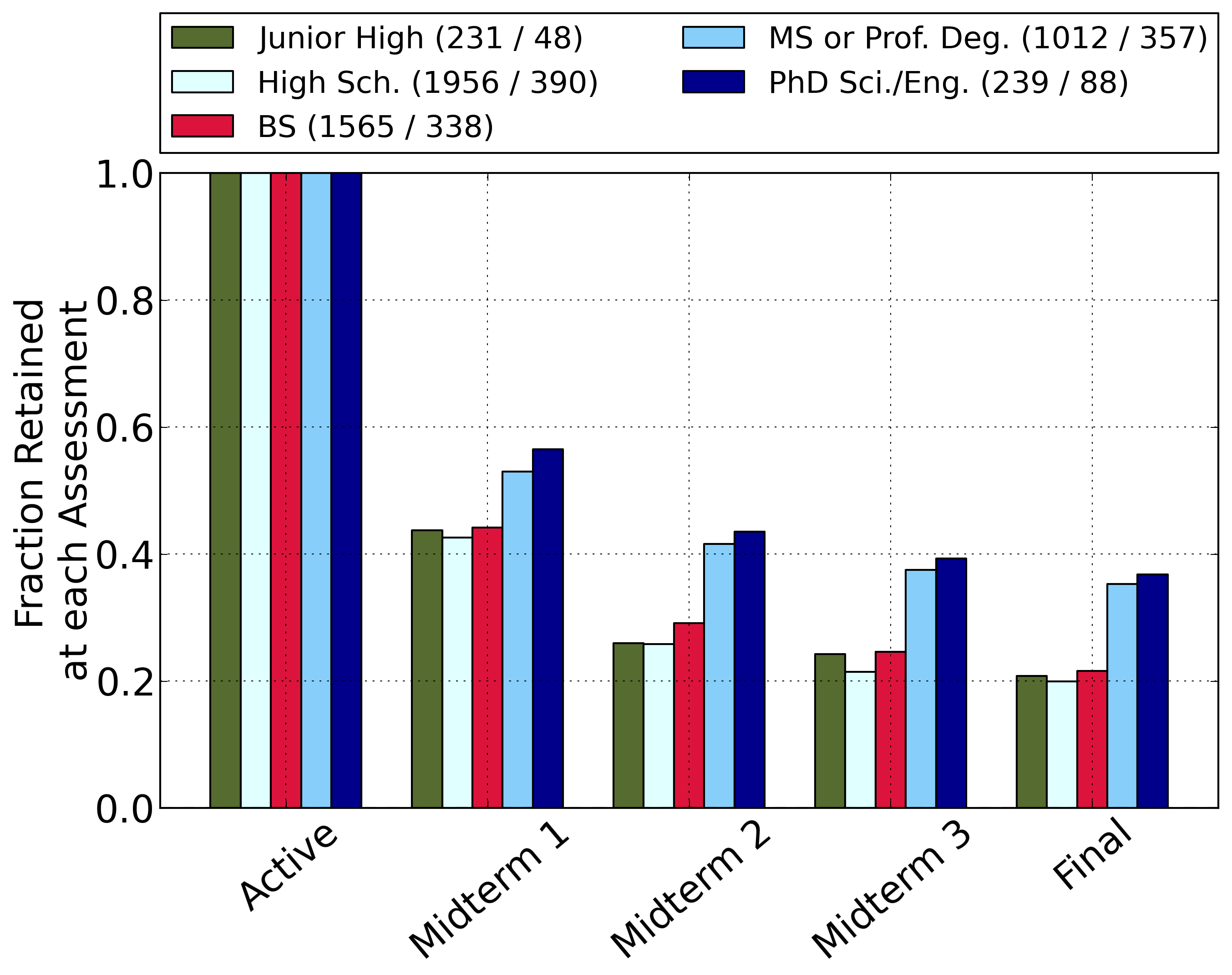}
\caption{\label{fig:all} 
Retention through participation for various levels of education: Junior High, High School, BS, MS/Prof. Deg., and PhD Sci./Eng. ``Fraction Retained'' represents the number of participants with score > 0 / number of "Active" participants (see methodology section). The ratio of Active to Final examinees is given in the legend (Active/Final). 
}
\end{figure}

\begin{figure}
\includegraphics[width=0.975\columnwidth]{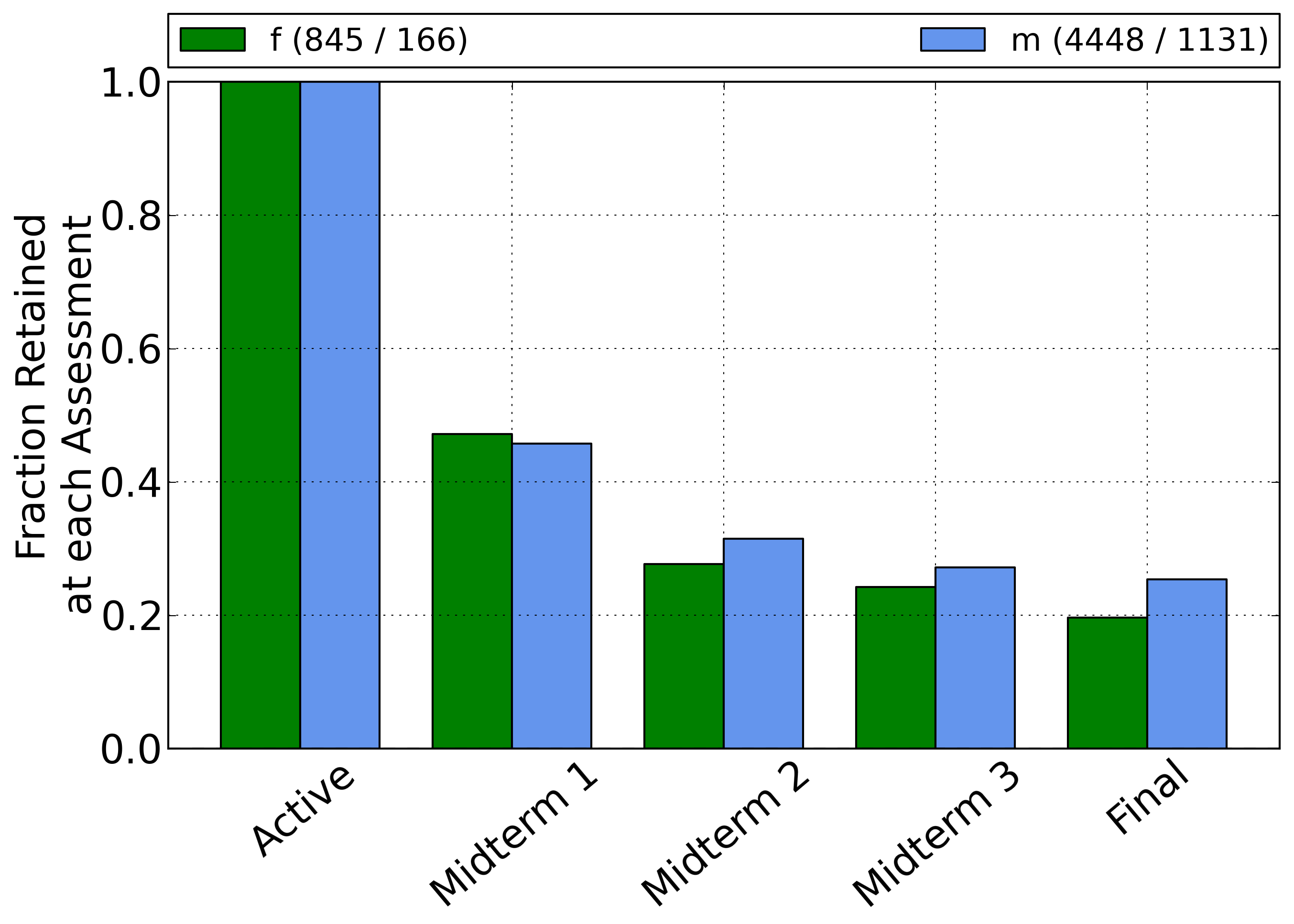}
\caption{\label{fig:gender} 
Retention through participation against gender. "Fraction Retained'' is measured with respect to "Active'' participants (see Fig.~\ref{fig:all}). The ratio of Active to Final examinees is given in the legend (Active/Final).
}
\end{figure}

\subsection{Who participated in 8.02x?} 
\hspace{0.19in}Retention in 8.02x has similar features to other studied MOOCs, namely a massive enrollment, with a relatively modest sized population of participants. Most MOOCs have less than 10\% of the initial enrollees who obtain certificates of completion\cite{MOOCretention}. Figure \ref{fig:retention} highlights the large attrition relative to enrollees and those taking the final exam. Retention rates with respect to "Active" students are more reasonable, but also subject to bias in our selection of "Active" students (students with > 500 events in weeks before the first midterm). 

Figure \ref{fig:all} presents retention versus level of education relative to the estimate of "Active" participants. Advanced degree holding participants are retained at nearly a 20\% difference compared to those with Bachelor's degrees or lower (note the size of each population is given as a ratio in the legend - Active / Final Examinees). The Junior High and PhD populations are perhaps too small for statistical inference, but differences between the three major groups imply that 8.02x might be better suited for advanced degree holders, i.e., they simply have more appropriate physics and math backgrounds. Equally interesting is to consider the possibility of advanced degree holders having better study and organization skills suitable for a self-regulated learning environment.

Figure \ref{fig:gender} presents a similar retention analysis for gender, highlighting a higher retention for males. Particularly interesting is the drop after the first midterm, as well as the slight shift in those taking the final. Further analysis is needed to investigate sources of the difference, possibly due to variation of age, education, or preparation level for the course.


\subsection{Performance and Level of Education}
\begin{figure}
\includegraphics[width=0.975\columnwidth]{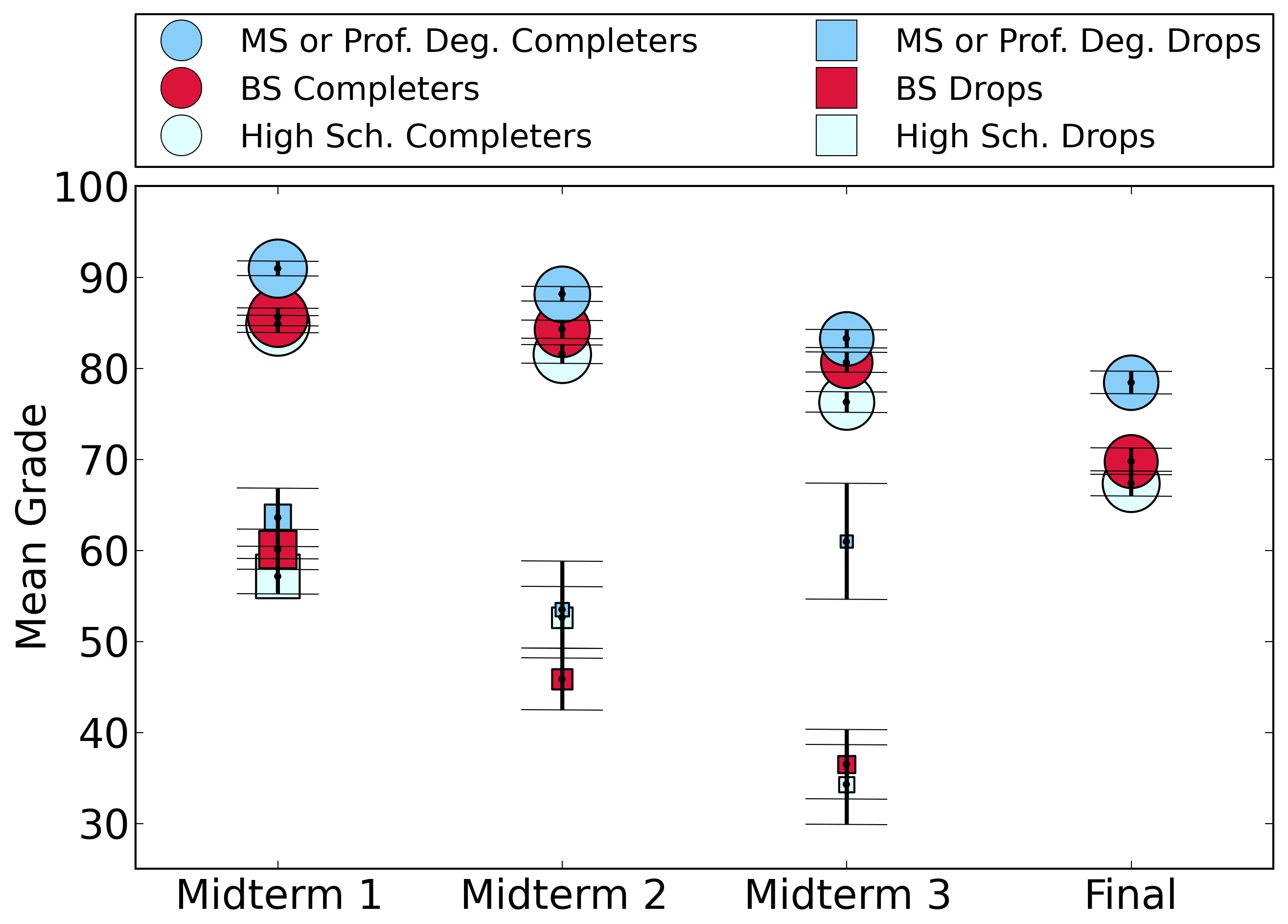}
\caption{\label{fig:allscore} 
Mean grade on examinations for varying level of education of participants; High School, BS, and MS / Prof. Deg. Each education level is divided into retained (circles) and drops (squares); drops are students that do not take the next exam (hence, no drops for final exam), and retained is based only on activity in the next exam (not the entire course). Shape size in both cases is relative to the number of participants in each category. Error bars are standard errors of the mean.
}
\end{figure}

\hspace{0.19in}Figure~\ref{fig:allscore} analyzes the examination performance of participants in the three largest leves of education categories. Each category is divided into "Completers'' - those who move on to the next exam, and "Drops'' - those who do not participate in the next exam (hence, no Drops for the Final). The mean grades of "Completers'' have relatively small differences (but statistically significant in most cases) with respect to level of education. There is also a noticeable drift to lower mean grade for all levels of education, either indicating increasing exam difficulty or student fatigue over the 16 course weeks, or both.  

The mean grades of "Drops'' in Fig.~\ref{fig:allscore} show that many people stop taking exams (a proxy for course activity) after earning a low grade. In future work, we plan to cross validate this signal by looking at the daily activity of "Drops'' after they have stopped taking exams. Some participants may have remained active after poor performance, but more detailed analysis of daily activity is required \cite{Kizilcec2013}.

\section{Discussion and Conclusions}
\hspace{0.19in}This study provides a first look at the first edX physics MOOC: 8.02x. Of particular importance are the results highlighting the incredibly diverse population of enrollees. 
Variation in age and level of education are beyond what typical physics instructors encounter in on-campus courses. Age differences between the top two countries (USA and India) are also intriguing, pointing toward deeper contexts in motivation for enrollment and participation. 

Our analysis of retention points toward the possible outcome of 8.02x being better suited for advanced degree holders versus college undergraduates. Some MOOCs have realized the importance of designing courses for population with education beyond a bachelor degree (e.g. a MOOC targeting high school physics teachers \cite{Fredericks2013}).

Additionally, differences in retention by gender provide the physics education research community insight into a new application of well-established gender research methodologies.       

Our results are aimed at highlighting both the challenges and opportunities associated with MOOC analysis.
Moving forward, we hope to bring more insight into behavioral patterns exhibited by learners in 8.02x, and future MOOC offerings. We plan to look deeper into the behavior of the different demographic groups, and investigate any differences in their learning associated with their habits. Furture research will focus on teasing out which components of the course, and which behavior patterns correlate with improved performance. 

\section{Acknowledgments}
\hspace{0.19in}We thank the Office of Digital Learning at MIT, and the Google Faculty Awards Program for their support of this project.

\bibliographystyle{abbrv}

\end{document}